\documentclass[twocolumn,english,prd,showpacs,showkeys,preprintnumbers,notitlepage]{revtex4}
\usepackage[letterpaper]{geometry}
\usepackage{amsmath}
\usepackage{graphicx}
\usepackage{amssymb}
\usepackage{esint}

\makeatletter

\def\Pom{{ I\!\!P}}\usepackage[hyperref]{hyperref}

\makeatother

\begin{document}

\title{Neutrino-production of single pions: dipole description}

\author{B.Z.~Kopeliovich}

\email{Boris.Kopeliovich@usm.cl}

\author{Ivan~Schmidt}

\email{Ivan.Schmidt@usm.cl}

\author{M.~Siddikov}

\email{Marat.Siddikov@usm.cl}

\affiliation{Departamento de F\'{\i}sica, Centro de Estudios Subat\'omicos, y Centro Cient\'ifico - Tecnol\'ogico de Valpara\'iso, Universidad T\'ecnica Federico Santa Mar\'{\i}a, Casilla 110-V, Valpara\'iso, Chile}
\begin{abstract}
The light-cone distribution amplitudes for the axial current are derived
within the instanton vacuum model (IVM), which incorporates nonperturbative
effects including spontaneous chiral symmetry breaking. This allows
to extend applicability of the dipole approach, usually used in the
perturbative domain, down to $Q^{2}\to0$, where partially conserved
axial current (PCAC) imposes a relation between the neutrino-production
cross section and the one induced by pions. A dramatic breakdown of
the Adler relation (AR) for diffractive neutrino-production of pions,
caused by absorptive corrections, was revealed recently in \cite{kpss}.
Indeed, comparing with the cross section predicted by the dipole phenomenology
at $Q^{2}\to0$ on a proton target we confirmed the sizable deviation
from the value given by the AR, as was estimated in \cite{kpss} within
a simplified two-channel model. The dipole approach also confirms
that in the black-disc limit, where the absorptive corrections maximize,
the diffractive cross section ceases, on the contrary to the expectation
based on PCAC. 
\end{abstract}



\preprint{USM-TH-288}

\maketitle

\section{Introduction }

Due to its $V$-$A$ shape the neutrino-hadron interactions possess
a very rich structure. However, because of the smallness of the cross-sections
until recently the experimental data have been scarce, mostly being
limited to the total cross-sections. With the launch of the new high-statistics
experiments like MINER$\nu$A at Fermilab~\cite{Drakoulakos:2004gn},
now the neutrino-hadron interactions may be studied with a better
precision. The $V$-$A$ structure of the neutrino-quark vertices
enables us to study simultaneously $\langle VV\rangle$, $\langle AA\rangle$
and $\langle VA\rangle$ correlators in the same process.

The properties of the vector current have been well studied in the
processes of deep inelastic scattering (DIS) of leptons on protons
and nuclei, deeply virtual Compton scattering (DVCS), real Compton
scattering (RCS), and vector meson production. The standard approach
in the description of such processes is based on the large-$Q^{2}$
factorization of the cross section into a process-dependent hard part,
which is evaluated in pQCD, and a universal target-dependent soft
part. The latter is extracted from fits to experimental data. Factorization,
however, is not valid at small photon virtualities, where one can
rely on the dispersion relation, or on the assumption of generalized
vector meson dominance (GVMD)~\cite{Fraas:1974gh,Ditsas:1975vd,Shaw:1993gx,Goeke:2008rn}.
Such a description, however, involves a lot of ad-hoc modeling.

An alternative phenomenology for high-energy QCD processes is based
on the color dipole approach~\cite{Kopeliovich:1981pz}. One assumes
that before interaction, the projectile (virtual $W$- or $Z$-boson
in case of the neutrino scattering) fluctuates into a quark-antiquark
dipole. After the dipole is formed, it scatters in the field of the
target and then fluctuates back to the final hadron~\cite{Kopeliovich:1981pz}.
Recently the color dipole approach has been successfully applied to
the description of different reactions with vector currents (see \cite{GolecBiernat:1998js,GolecBiernat:2004xu,Hufner:2000jb,Kopeliovich:2008ct,Kopeliovich:2009cx,Kopeliovich:2010sa,Kopeliovich:2010xm,Fiore:2005bp,Fiore:2005yi,Fiore:2008cc,Fiore:2008nj,Floter:2007xv,GayDucati:2008hi,GayDucati:2008zzc,Ducati:2006vh,Machado:2007wq,Machado:2008tp,Machado:2008zv,Machado:2009cd,McFarland:2006pz}
and references therein).

For the axial current the situation is more complicated, especially
at small $Q^{2}$, because the chiral symmetry breaking generates
the near-massless pseudo-goldstone mesons (pions). Straightforward
extension of the vector dominance model to the axial current leads
to the so called Piketty-Stodolsky paradox~\cite{Piketty:1970sq,Kopeliovich:1992ym},
which appears because axial meson dominance is broken by a large contribution
of multi-pion singularities in the dispersion relation ~\cite{Belkov:1986hn,Kopeliovich:1992ym}.
The dipole description is free of this problem, because in this model
there is no explicit hadronic degrees of freedom and the interaction
occurs via dipole scattering.

According to the Adler theorem~\cite{Adler:1964yx,Adler:1966gc},
based on the hypothesis of partial conservation of axial current (PCAC),
the neutrino-proton interactions cross-section at zero $Q^{2}$ is
proportional to the cross-section of the pure hadronic process, where
the heavy intermediate boson is replaced by a pion, \begin{equation}
\left.\nu\,\frac{d\sigma_{\nu p\to lF}}{d\nu\, dQ^{2}}\right|_{Q^{2}=0}=\frac{G_{F}^{2}}{2\pi^{2}}\, f_{\pi}^{2}\,(1-y)\,\sigma_{\pi p\to F}(\nu),\label{eq:Adler}\end{equation}
 where $F$ denotes the final hadronic state; $y=\nu/E_{\nu}$; $E_{\nu}$
is the energy of the neutrino; and $\nu$ is the energy of the heavy
boson $W$, or $Z$, in the target rest frame. The chiral symmetry
is vital and should be embedded into any dynamical model which is
used for calculation of the cross section at small $Q^{2}$.

In what follows we entirely neglect the lepton mass (accurate for
neutral currents or electrons), which can be easily incorporated~\cite{Piketty:1970sq},
but is dropped for the sake of simplicity. If one interpreted the
Adler relation (AR) (Eq.~\ref{eq:Adler}) in terms of pion pole dominance,
one would arrive at a vanishing cross section. Indeed, the pion pole
term in the amplitude contains a factor $q_{\mu}$, which multiplied
by the conserved lepton current $l_{\mu}$ terminates this contribution~\cite{Bell:1978qu,Piketty:1970sq,Kopeliovich:1992ym,kpss}.
Other, heavier meson states provide a final contribution, but have
to conspire to mimic the pion pole term, as is dictated by the PCAC
relation Eq.~(\ref{eq:Adler}).

Such a fine tuning looks miraculous if one has no clue of the underlying
dynamics. Similar paradox is known for the $1/Q^{2}$ behavior of
DIS cross section. In QCD this is known as a result of color screening
\cite{Kopeliovich:1981pz}, leading to the effect called color transparency.
While this is rather obvious within the color dipole description,
it becomes extremely sophisticated in hadronic representation. Indeed,
expanding the current over hadronic states one may arrive at a problem
called Bjorken paradox~\cite{bjorken}. Namely, how it happens that
many hadronic states, having large sizes and large cross sections
conspire in a way that all together they act like a small hadron with
a tiny, $\sim1/Q^{2}$ cross section. The solution is known, the off
diagonal diffractive amplitudes are negative and cancel with diagonal
ones \cite{mukhin}. This however cannot be proven, unless one employs
an explicit model describing the features of the hadronic states and
the diffractive amplitudes. This is why the color transparency effect
has not been understood within the GVMD, but was revealed in the color
dipole representation \cite{Kopeliovich:1981pz}.

Similarly, in order to test the mysterious relation between the contribution
of heavy hadronic fluctuations and pion, one should switch to the
dipole representation and employ models for the distribution amplitudes
(DA) of the axial current which have built-in chiral symmetry. Recently,
we used the DA of the vector current calculated in the IVM for the
evaluation of several processes~\cite{Kopeliovich:2008ct,Kopeliovich:2009cx,Kopeliovich:2010xm,Kopeliovich:2010sa}.
In this paper we apply the IVM to construct the DAs for the axial
current and pion and use them to calculate the neutrino-production
cross sections. Since the IVM includes spontaneous chiral symmetry
breaking, the $\bar{q}q$ DAs of axial current and pion should automatically
satisfy PCAC, and in the small-$Q^{2}$ limit reproduce the Adler
relation~(\ref{eq:Adler}).

Notice that the color dipole description is valid only at high-energies
or at small $x_{B}\ll1$, where the contribution of quark exchange
(reggeons) is suppressed as $1/\sqrt{\nu}$. For moderate energies
another mechanisms, such as e.g. formation of resonances in the direct
channel~\cite{Lalakulich:2006sw,Lalakulich:2006yn}, and/or reggeon exchange in the crossed channel may be important~\cite{Gershtein:1980vd,Komachenko:1983jv}.
In this paper we do not consider those corrections, but concentrate
on the well developed small-x dipole phenomenology.

Experimentally, the neutrino-production of hadrons on protons and
nuclei has been studied in the recent experiments K2K~\cite{Ahn:2002up,Hasegawa:2005td,Gran:2006jn},
MiniBoone~\cite{:2007ru,AguilarArevalo:2008xs}, NuTeV~\cite{Zeller:2001hh,Goncharov:2001qe,Romosan:1996nh}
(see also review~\cite{Kopeliovich:1992ym} and 
\cite{Belkov:1986hn,Amsler:2008zzb,Kopeliovich:1990kd,Kopeliovich:1989wc,Kopeliovich:2004px}
for references to earlier neutrino experiments). For high energy neutrino
scattering, there are data from the early bubble chamber experiments~\cite{Bell:1978qu,Allen:1985ti}
with energies up to 100 GeV, though with low statistics and only for
the total (integrated) cross-sections. Currently, with the launch
of the high statistics experiment Minerva at Fermilab~\cite{Drakoulakos:2004gn,McFarland:2006pz},
the precision of measurements should be considerably improved, and
data for the differential cross-sections at high energies will become
available.

In this paper we consider a particular process -- diffractive single
pion production on a proton target. As was demonstrated in \cite{kpss}
and below, this process provides a most sensitive way to test PCAC
in high energy neutrino interactions. Besides, it generates an important
background to the measurements of neutrino oscillations~\cite{Ashie:2005ik,Ahmad:2002jz,Ahn:2006zza,Eguchi:2002dm},
and is also important for the neutrino astronomy of astrophysical
and cosmological sources.

The paper is organized as follows. In Section~\ref{sec:proton} we
present the color dipole formalism. In Section~\ref{sec:WFfromIVM}
we perform calculations of the DAs of the axial current and pion.
In Section~\ref{sec:Overlap-details} we calculate the overlap of
the DAs for the axial current and of the pion and found it to be proportional
to $q_{\mu}$, what terminates this contribution to the neutrino-production
of pions due to conservation of lepton current. In Section~\ref{sec:Results}
we present the numerical results and summarize the observations in
Section~\ref{summary}.

\section{Diffractive production of pions}

\label{sec:proton}

The cross section of diffractive neutrino-production of a pion on
a proton, $\nu p\to l\pi p$, has the form, \begin{equation}
\nu\,\frac{d^{3}\sigma_{\nu p\to\mu\pi p}}{d\nu dtdQ^{2}}=\frac{G_{F}^{2}\ L_{\mu\nu}\left(W_{\mu}^{A}\right)^{*}W_{\nu}^{A}}{32\pi^{3}m_{N}^{2}E_{\nu}^{2}\sqrt{1+Q^{2}/\nu^{2}}},\label{eq:disgma_XSection}\end{equation}
 where $m_{N}$ is the nucleon mass; $L_{\mu\nu}$ is the lepton tensor;
and $W_{\mu}^{A}$ is the amplitude of pion production by the axial
current on the proton target. In the color dipole model this amplitude
has the form \begin{eqnarray}
W_{\mu}^{A}\left(s,\Delta,Q^{2}\right) & = & \int\limits _{0}^{1}d\beta_{1}d\beta_{2}d^{2}r_{1}d^{2}r_{2}\bar{\Psi}^{\pi}\left(\beta_{2},\vec{r}_{2}\right)\label{eq:W_amplitude_defintiion}\\
 & \times & \mathcal{A}^{d}\left(\beta_{1},\vec{r}_{1};\beta_{2},\vec{r}_{2};\Delta\right)\Psi_{\mu}^{A}\left(\beta_{1},\vec{r}_{1}\right)\nonumber \end{eqnarray}
 where $\bar{\Psi}^{\pi}$ and, $\Psi_{\mu}^{A}$ are the distribution
amplitudes (DAs) of the pion and axial current respectively, and $\mathcal{A}^{d}(...)$
is the dipole scattering amplitude. The axial current DA $\Psi_{\mu}^{A}$
contains a pion pole, whose contribution to the amplitude is proportional
to $q_{\mu}$, because the pion is spinless. This factor terminates
the pion pole because of conservation of the lepton current. As we
assumed, the lepton is massless, otherwise the pion pole contribution
is not zero and leads to corrections of the order of $\mathcal{O}\left(\frac{m_{l}^{2}}{m_{\pi}^{2}+Q^{2}}\right)$
\cite{Piketty:1970sq}.

The amplitude $\mathcal{A}^{d}\left(\beta_{1},\vec{r}_{1};\beta_{2},\vec{r}_{2};\Delta\right)$
in (\ref{eq:W_amplitude_defintiion}) depends on the initial and final
quark transverse separations $\vec{r}_{1,2}$, fractional light-cone
momenta $\beta_{1,2}$, and transverse momentum transfer $\vec{\Delta}$.
This is a universal function dependent only on the target but not
on the initial and final states. In addition to the axial current
contribution, in~(\ref{eq:W_amplitude_defintiion}) there should
be the contribution of the vector current. This contribution involves
a poorly known helicity flip dipole amplitude $\tilde{\mathcal{A}}_{d}$,
which vanishes at high energies as $1/\nu$. Besides, at small $Q^{2}$,
the vector current contribution is suppressed by a factor $Q^{2}$.
At high energies, in the small angle approximation, $\Delta/\sqrt{s}\ll1$,
and the quark separation and fractional momenta $\beta$ are preserved,
so \begin{eqnarray}
\mathcal{A}^{d}\left(\beta_{1},\vec{r}_{1};\beta_{2},\vec{r}_{2};Q^{2},x,\Delta\right) & \approx & \delta\left(\beta_{1}-\beta_{2}\right)\delta\left(\vec{r}_{1}-\vec{r}_{2}\right)\label{eq:DVCSIm-BK}\\
 & \times & (\epsilon+i)\Im m\, f_{\bar{q}q}^{N}\left(\vec{r},\vec{\Delta},\beta,x\right)\nonumber \end{eqnarray}
 where $\epsilon$ is the ratio of the real to imaginary parts, and
for the imaginary part of the elastic dipole amplitude we employ the
model developed in \cite{Kopeliovich:2007fv,Kopeliovich:2008nx,Kopeliovich:2008da,Kopeliovich:2008ct},
\begin{widetext} \begin{eqnarray}
\Im m\, f_{\bar{q}q}^{N}\left(\vec{r},\vec{\Delta},\beta,x\right)=\frac{\sigma_{0}(x)}{4}\exp\left[-\left(\frac{B(x)}{2}+\frac{R_{0}^{2}(x)}{16}\right)\vec{\Delta}_{\perp}^{2}\right]\left(e^{-i\beta\vec{r}\cdot\vec{\Delta}}+e^{i(1-\beta)\vec{r}\cdot\vec{\Delta}}-2e^{i\left(\frac{1}{2}-\beta\right)\vec{r}\cdot\vec{\Delta}}e^{-\frac{r^{2}}{R_{0}^{2}(x)}}\right).\label{eq:ImF}\end{eqnarray}
 \end{widetext} The phenomenological functions $\sigma_{0}(x)$,
$R_{0}^{2}(x)$ and $B(x)$ are fitted to DIS and $\rho$ electroproduction
data. We rely here on the Bjorken variable $x=Q^{2}/2(pq)$, which
has the meaning of fractional light-cone momentum of the parton only
at large $Q^{2}$. At low $Q^{2}$ important for the axial current,
one should switch to an energy dependent parametrization, as is explained
in Section~\ref{sec:Results}.

For the forward scattering, $\Delta\to0$, the imaginary part of the
amplitude (\ref{eq:DVCSIm-BK}) reduces to the saturated parameterization
of the dipole cross-section proposed by Golec-Biernat and W\"usthoff
(GBW)~\cite{GolecBiernat:1998js}, \begin{eqnarray}
\sigma_{d}(r,x) & = & \Im m\, f_{\bar{q}q}^{N}\left(\vec{r},\vec{\Delta}=0,\beta,x\right)\label{eq:Sigma_QQ_GPD}\\
 & = & \sigma_{0}(x)\left[1-\exp\left(-\frac{r^{2}}{R_{0}^{2}(x)}\right)\right].\nonumber \end{eqnarray}
 Generally speaking, the amplitude $f_{\bar{q}q}^{N}(...)$ involves
nonperturbative physics, but its asymptotic behavior at small~$r$
is controlled by pQCD \cite{Kopeliovich:1981pz}:\[
f_{\bar{q}q}^{N}({r})_{r\to0}\propto r^{2},\]
 up to slowly varying factors $\sim\ln(r)$ \cite{Kopeliovich:1981pz}.

Calculation of the differential cross section also involves the real
part of scattering amplitude, whose relation to the imaginary part
is quite straightforward. According to \cite{Bronzan:1974jh}, if
$\lim\limits _{s\to\infty}\left(\frac{\mathcal{I}m\,{f}}{s^{\alpha}}\right)$
is finite, then the real and imaginary parts of the forward amplitude
are related as\begin{equation}
\mathcal{R}e\,{f(\Delta=0)}=s^{\alpha}\tan\left[\frac{\pi}{2}\left(\alpha-1+\frac{\partial}{\partial\ln s}\right)\right]\frac{\Im m\,{f(\Delta=0)}}{s^{\alpha}}.\label{eq:BronzanFul}\end{equation}
 In the model under consideration the imaginary part of the forward
dipole amplitude indeed has a power dependence on energy, $\mathcal{I}m\, f(\Delta=0;\, s)\sim s^{\alpha_{\Pom}-1}$,
where $\alpha_{\Pom}$ is the intercept of the effective Pomeron trajectory.
Then Eq.~(\ref{eq:BronzanFul}) simplifies to \begin{eqnarray}
\frac{\mathcal{R}e\,\mathcal{A}}{\Im m\,\mathcal{A}} & =\tan\left(\frac{\pi}{2}(\alpha_{\Pom}-1)\right)\equiv\epsilon.\end{eqnarray}

This fixes the phase of the forward scattering amplitude, which we
retain for nonzero momentum transfer, assuming similar $\Delta$-dependences
for the real and imaginary parts.

\section{Distribution amplitudes and the instanton vacuum model}

\label{sec:WFfromIVM}In this section we define the DAs and give a
brief description of the instanton model used for their evaluation~~(see~\cite{Schafer:1996wv,Diakonov:1985eg,Diakonov:1995qy}
and references therein).

\subsection{Instanton vacuum model}

\label{ivm}

The central object of the model is the effective action for light
quarks in the instanton vacuum, which in the leading order in $N_{c}$
has the form~\cite{Diakonov:1985eg,Diakonov:1995qy} \begin{eqnarray}
S & = & \int d^{4}x\left(\frac{N}{V}\ln\lambda+2\Phi^{2}(x)\right.\label{eq:effact}\\
 & - & \left.\bar{\psi}\left(\hat{p}+\hat{v}+\hat{a}\gamma_{5}-m-c\bar{L}f\otimes\Phi\cdot\Gamma_{m}\otimes fL\right)\psi\right)\nonumber \end{eqnarray}
 where $\Gamma_{m}$ is one of the matrices, $\Gamma_{m}=1,i\vec{\tau},\gamma_{5},i\vec{\tau}\gamma_{5}$;
$\psi$ and $\Phi$ are the fields of constituent quarks and mesons
respectively; $N/V$ is the density of the instanton gas; $m\approx5$~MeV
is the current quark mass; $\hat{v}\equiv v_{\mu}\gamma^{\mu}$ is
the external vector current corresponding to the photon. $L$ is the
gauge factor defined as, \begin{eqnarray}
L\left(x,z\right) & = & P\exp\left(i\int\limits _{z}^{x}d\zeta^{\mu}\left(v_{\mu}(\zeta)+a_{\mu}\left(\zeta\right)\gamma_{5}\right)\right).\label{eq:L-factor}\\
\bar{L}(x,z) & = & \gamma_{0}L^{\dagger}(x,z)\gamma_{0}\end{eqnarray}
 It provides the gauge invariance of the action, and $f(p)$ is the
Fourier transform of the zero-mode profile in the single-instanton
background. In this paper we used for evaluations the dipole-type
parameterization~\cite{Diakonov:1985eg}\[
f(p)=\frac{L^{2}}{L^{2}-p^{2}}\]
 with $L\sim850\,$MeV.

In the leading order in $N_{c}$, we have the same Feynman rules as
in the perturbative theory, but with a momentum-dependent quark mass
$\mu(p)$ in the quark propagator\begin{eqnarray}
S(p) & = & \frac{1}{\hat{p}-\mu(p)+i0}.\end{eqnarray}
 The mass of the constituent quark has the form \[
\mu(p)=m+M\, f^{2}(p),\]
 where $m\approx5$~MeV is the current quark mass, $M\approx350$~MeV
is the dynamical mass generated by the interaction with the instanton
vacuum background. Due to the presence of the instantons the vector
current - quark coupling is also modified,\begin{eqnarray*}
\hat{v} & \equiv & v_{\mu}\gamma^{\mu}\rightarrow\hat{V}=\hat{v}+\hat{V}^{nonl},\\
\hat{a} & \equiv & a_{\mu}\gamma^{\mu}\rightarrow\hat{A}=\hat{a}+\hat{A}^{nonl},\end{eqnarray*}
 In addition to the vertices of the perturbative QCD, the model contains
the nonlocal terms with higher-order couplings of currents to mesons.
The exact expressions for the nonlocal terms $\hat{V}^{nonl},\hat{A}^{nonl}$
depend on the choice of the path in~(\ref{eq:L-factor}), so one
can find different results in the literature~\cite{Dorokhov:2006qm,Anikin:2000rq,Dorokhov:2003kf,Goeke:2007j}.
However, for the longitudinal parts of the axial and vector currents
important here, this ambiguity cancels out and couplings have the
form\begin{eqnarray*}
\hat{V}_{nonl} & = & v_{\mu}\left[iM\frac{p_{1}^{\mu}+p_{2}^{\mu}}{p_{2}^{2}-p_{1}^{2}}\left(f\left(p_{2}\right)^{2}-f\left(p_{1}\right)^{2}\right)\right],\\
\hat{A}_{nonl} & = & a_{\mu}\left[iM\frac{p_{1}^{\mu}+p_{2}^{\mu}}{p_{2}^{2}-p_{1}^{2}}\Bigl(f\left(p_{2}\right)-f\left(p_{1}\right)\Bigr)^{2}\right],\end{eqnarray*}
 where $p_{1},\, p_{2}$ are the momenta of the initial and final
quarks.

\subsection{Axial current distribution amplitudes}

\label{axialDA}

\begin{figure}[htb]
 \includegraphics[scale=0.55]{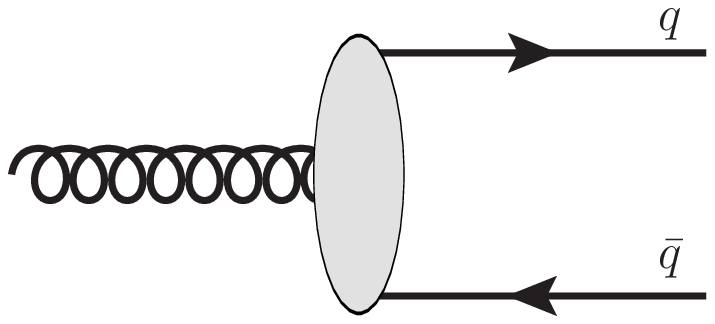}\includegraphics[scale=0.55]{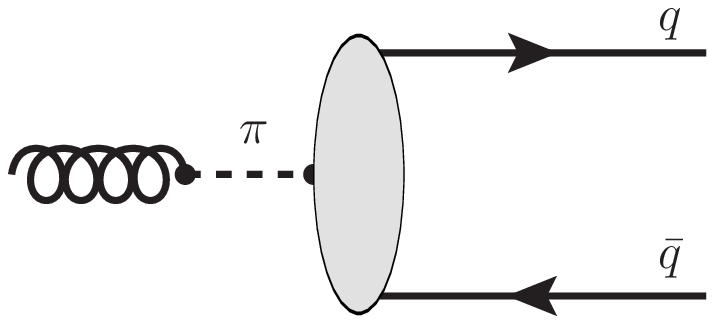}
\caption{\label{fig:DA_12}The distribution amplitude has two contributions,
with intermediate heavy axial states and with a pion, labeled with
\textit{(bulk)} and \textit{(pion)} respectively.}

\end{figure}

The distribution amplitudes of the axial current are defined via 3-point
correlators \begin{equation}
\Psi_{\beta}\sim\int d^{4}\xi\, e^{-iq\cdot\xi}\left\langle 0\left|\bar{\psi}\left(y\right)\Gamma\psi\left(x\right)J_{\beta}^{5}(\xi)\right|0\right\rangle ,\label{eq:Psi_mu}\end{equation}
 where $J_{\beta}^{5}(\xi)$ is the axial isovector current and $\Gamma$
is one of the Dirac matrices. Due to the spontaneous chiral symmetry
breaking and existence of the near-massless pions the hadronic structure
of the axial current differs from that of the vector current. In particular,
the axial current may fluctuate into the pion state before production
of the $\bar{q}q$ pair. Thus the correlator~(\ref{eq:Psi_mu}) has
two contributions, schematically shown in the Figure~\ref{fig:DA_12}.

One term corresponds to the combined contribution of the intermediate
heavy states ($a_{1}$ meson, $3\pi$, etc.), another one corresponds
to the axial current fluctuating into a pion. Due to the built-in
chiral symmetry, the two contributions are connected by PCAC, so for
the full DA we have \begin{equation}
\Psi_{\mu}=\Psi_{\mu}^{(bulk)}+\Psi_{\mu}^{(pion)}=\left(g_{\mu\nu}-\frac{q_{\mu}q_{\nu}}{q^{2}-m_{\pi}^{2}}\right)\Psi_{\nu}^{(bulk)}.\label{eq:Psi_nonsing}\end{equation}
 This form of the DA reflects the relation between the pion pole and
the bulk of heavy states contribution imposed by PCAC, which has been
discussed above. In what follows we concentrate on the part of the
amplitude presented in the dispersion relation by the bulk of heavy
states excluding the pion pole (left pane in the Figure~\ref{fig:DA_12}),
tacitly assuming that the full distribution amplitudes are determined
using (\ref{eq:Psi_nonsing}).

For the part of the axial current presented by the bulk of heavy states,
the DAs may be defined similar to the distribution amplitudes of the
axial meson~\cite{Yang:2007zt}:\begin{widetext}\begin{eqnarray}
\left\langle 0\left|\bar{\psi}\left(y\right)\gamma_{\mu}\gamma_{5}\psi\left(x\right)\right|A(q)\right\rangle  & = & if_{A}^{2}\int_{0}^{1}d\beta\, e^{i(\beta p\cdot y+\bar{\beta}p\cdot x)}\nonumber \\
 & \times & \left[p_{\mu}\frac{e^{(\lambda)}\cdot z}{p\cdot z}\Phi_{||}(\beta)+e_{\mu}^{(\lambda=\perp)}g_{\perp}^{(a)}(\beta)-\frac{1}{2}z_{\mu}\frac{e^{(\lambda)}\cdot z}{(p\cdot z)^{2}}f_{A}^{2}g_{3}(\beta)\right],\label{eq:AWF-mu5}\end{eqnarray}

\begin{eqnarray}
\left\langle 0\left|\bar{\psi}\left(y\right)\gamma_{\mu}\psi\left(x\right)\right|A(q)\right\rangle  & = & -if_{A}^{2}\epsilon_{\mu\nu\rho\sigma}e_{\nu}^{(\lambda)}p_{\rho}z_{\sigma}\int_{0}^{1}d\beta\, e^{i(\beta p\cdot y+\bar{\beta}p\cdot x)}\frac{g_{\perp}^{(v)}(\beta)}{4}\label{eq:AWF-mu}\end{eqnarray}

\begin{eqnarray}
\left\langle 0\left|\bar{\psi}\left(y\right)\sigma_{\mu\nu}\gamma_{5}\psi\left(x\right)\right|A(q)\right\rangle  & = & f_{A}\int_{0}^{1}d\beta\, e^{i(\beta p\cdot y+\bar{\beta}p\cdot x)}\biggl[\left(e_{\mu}^{(\lambda=\perp)}p_{\nu}-e_{\nu}^{(\lambda=\perp)}p_{\mu}\right)\Phi_{\perp}(\beta)\nonumber \\
 & + & \left.\frac{e^{(\lambda)}\cdot z}{(p\cdot z)^{2}}f_{A}^{2}\left(p_{\mu}z_{\nu}-p_{\nu}z_{\mu}\right)h_{||}^{(t)}(\beta)+\frac{1}{2}\left(e_{\mu}^{(\lambda)}z_{\nu}-e_{\nu}^{(\lambda)}z_{\mu}\right)\frac{f_{A}^{2}}{p\cdot z}h_{3}(\beta)\right],\label{eq:AWF-munu}\end{eqnarray}

\begin{eqnarray}
\left\langle 0\left|\bar{\psi}\left(y\right)\gamma_{5}\psi\left(x\right)\right|A(q)\right\rangle  & = & f_{A}^{3}e^{(\lambda)}\cdot z\int_{0}^{1}d\beta\, e^{i(\beta p\cdot y+\bar{\beta}p\cdot x)}\frac{h_{||}^{(p)}(\beta)}{2},\label{eq:AWF-5}\end{eqnarray}
 \end{widetext} where $q$ is the momentum carried by the axial current;
$\beta$ is the fractional light-cone momentum; $\bar{\beta}\equiv1-\beta$,
$e^{(\lambda)}\equiv e^{(\lambda)}(q)$ is the polarization vector
of the axial meson with polarization state $\lambda$; $z=x-y$; $p_{\mu}$
is the {}``positive direction'' vector on the light-cone; $n_{\mu}$
is the {}``negative direction'' vector on the light-cone. Light
cone vectors $p,n$ are chosen in such a way that the vector $q$
does not have transverse components. The normalization constant $f_{A}$
is a dimensional parameter introduced in order to make the distribution
amplitudes dimensionless. Its value is fixed from the condition\begin{equation}
\int_{0}^{1}d\beta\,\Phi_{||}(\beta)=1.\label{eq:fANorm}\end{equation}
 We defined an {}``effective'' axial state $\left|A^{(\lambda)}(q)\right\rangle $
as\begin{equation}
\left|A^{(\lambda)}(q)\right\rangle =\int d^{4}x\, e^{-iq\cdot x}e_{\beta}^{(\lambda)}(q)J_{\beta}^{5}(x)\left|0\right\rangle .\label{eq:A-defiition}\end{equation}
 The DAs have the following twists: $\Phi_{||}(\beta),\Phi_{\perp}(\beta)$
are twist-2; $g_{\perp}^{(a)},g_{\perp}^{(v)},h_{||}^{(t)},h_{||}^{(p)}$
are of twist-3; $g_{3},h_{3}$ are of twist-4. All the wave functions
in~(\ref{eq:AWF-mu5},\ref{eq:AWF-mu}) are chiral even; all the
wave functions in~(\ref{eq:AWF-munu},\ref{eq:AWF-5}) are chiral
odd.

\subsection{Pion distribution amplitudes}

\label{pion}

A spinless pion has only four independent DAs defined as~\cite{Ball:2006wn}:\begin{eqnarray}
\left\langle 0\left|\bar{\psi}\left(y\right)\gamma_{\mu}\gamma_{5}\psi\left(x\right)\right|\pi(q)\right\rangle  & = & if_{\pi}\int_{0}^{1}d\beta\, e^{i(\beta p\cdot y+\bar{\beta}p\cdot x)}\label{eq:piWF-mu5}\\
 & \times & \left(p_{\mu}\phi_{2;\pi}(\beta)+\frac{1}{2}\frac{z_{\mu}}{(p\cdot z)}\psi_{4;\pi}(\beta)\right),\nonumber \end{eqnarray}

\begin{eqnarray}
\left\langle 0\left|\bar{\psi}\left(y\right)\gamma_{5}\psi\left(x\right)\right|\pi(q)\right\rangle  & = & -if_{\pi}\frac{m_{\pi}^{2}}{m_{u}+m_{d}}\label{eq:piWF-5}\\
 & \times & \int_{0}^{1}d\beta\, e^{i(\beta p\cdot y+\bar{\beta}p\cdot x)}\phi_{3;\pi}^{(p)}(\beta),\nonumber \end{eqnarray}

\begin{eqnarray}
\left\langle 0\left|\bar{\psi}\left(y\right)\sigma_{\mu\nu}\gamma_{5}\psi\left(x\right)\right|\pi(q)\right\rangle  & = & -\frac{i}{3}f_{\pi}\frac{m_{\pi}^{2}}{m_{u}+m_{d}}\int_{0}^{1}d\beta\,\phi_{3;\pi}^{(\sigma)}(\beta)\nonumber \\
 & \times & \frac{e^{i(\beta p\cdot y+\bar{\beta}p\cdot x)}}{p\cdot z}\left(p_{\mu}z_{\nu}-p_{\nu}z_{\mu}\right),\label{eq:piWF-munu}\end{eqnarray}
 Twist counting is the following: $\phi_{2;\pi}$ is a single twist-2
function (it was evaluated earlier in \cite{Petrov:1998kg,Dorokhov:2006xw,Dorokhov:2002iu}),
$\phi_{3;\pi}^{(p)}$ and $\phi_{3;\pi}^{(\sigma)}$ are twist-3,
$\psi_{4;\pi}$ is the twist-4 DA.

The full expressions for the DAs (\ref{eq:AWF-mu5})-(\ref{eq:AWF-5})
and (\ref{eq:piWF-mu5})-(\ref{eq:piWF-munu}) are given in Appendixes~\ref{sec:axialDA}
and \ref{pionDA} respectively.

The DAs for the vector current are presented in Appendix~\ref{vectorDA}.
However, the vector current does not contribute to the color dipole
amplitudes of pion production. Although it contains nonzero components,
their overlap with the pion DAs is zero. The vector part vanishes
because the color dipole amplitude does not flip helicity. Within
the VDM approximation such components may be expressed via the $\rho N\to\pi N$
scattering amplitudes, which exist only due to quark-antiquark (Reggeon)
exchange in the cross channel. This is beyond the employed dipole
phenomenology corresponding to gluonic (Pomeron) exchanges.

\section{Disappearance of the pion pole in the dipole representation}

\label{sec:Overlap-details}

As was emphasized above, the pion pole contribution to the pion production
amplitude vanishes because of lepton current conservation (up to the
lepton mass). This nontrivial observation of \cite{Bell:1978qu,Piketty:1970sq,Kopeliovich:1992ym,kpss}
is in variance with the naive interpretation of the AR Eq.~(\ref{eq:Adler}),
which relates diffractive neutrino-production of pions, $\nu+p\to l+\pi+p$,
with elastic pion-proton scattering, $\pi+p\to\pi+p$. It is tempting
to interpret this relation as pion pole dominance, i.e. neutrino fluctuates
to a pion, which then interacts elastically with the proton target.
If this were true, the amplitude should be maximized in the so called
black disk limit, which correspond to unitarity saturation when the
imaginary part of the partial elastic amplitude reaches the maximal
value allowed by the unitarity relation.

On the other hand, if the pion pole does not contribute \cite{Bell:1978qu,Piketty:1970sq,Kopeliovich:1992ym,kpss}
as is stressed above, all hadronic fluctuations of the neutrino contributing
to $\Psi^{(bulk)}$ are heavier than a pion, so all diffractive hadronic
amplitudes of pion production are off-diagonal. Such amplitudes vanish
in the black-disc limit, so the pion cannot be produced diffractively.
The source of such a dramatic breakdown of PCAC was identified in
\cite{kpss} as a result of strong absorptive corrections. Of course
the deviation from the PCAC prediction, AR, on a proton or nuclear
targets, which may be far from the unitarity bound, is not so dramatic,
as was calculated in \cite{kpss}.

In this section we present an explicit demonstration of disappearance
of diffractive pion production in the black-disc limit relying on
the dipole model. Namely in this regime all the partial elastic amplitudes
(\ref{eq:ImF}) reach the unitarity bound, so become independent of
the dipole transverse separation $\vec{r}$, and the equation (\ref{eq:W_amplitude_defintiion})
simplifies to just an overlap of the initial (axial current) and final
(pion) light-cone wave function. We intent to demonstrate that this
overlap vanishes.

The amplitude of pion production in this regime has the form, \begin{eqnarray}
F_{\mu}^{J_{A}\to\pi}\left(q,\Delta\right) & = & \sum_{a}\int d\beta\, d^{2}r\nonumber \\
 & \times & \bar{\Psi}_{\pi}^{(a)}\left(\beta,\vec{r};q-\Delta\right)\Psi_{\mu,A}^{(a)}\left(\beta,\vec{r};q\right),\label{eq:App_Overlap_def}\end{eqnarray}
 where the index $a$ numerates all the distribution amplitudes. This
is suppressed, since transition from spin-1 to spin-0 requires helicity
flip for one of the quarks in the quark-antiquark pair. Now we would
like to demonstrate explicitly that such suppression indeed takes
place in case of the perturbative QCD model.

The distribution amplitude of the meson state is defined as\begin{equation}
\int dz\, e^{i\beta z}e^{i\vec{k}\vec{r}}\left\langle 0\left|\bar{\psi}(z,\vec{r})\Gamma\psi(0)\right|M(q)\right\rangle =\Phi_{M}\left(\beta,\vec{r};q\right),\label{eq:DA_definition}\end{equation}
 where the separation between the quark and antiquark has a {}``minus''
and transverse components $(z,r)$, and $\Gamma$ is one of the Dirac
matrices $\left(1,\gamma^{5},\gamma_{\mu},\gamma_{\mu}\gamma_{5},\sigma_{\mu\nu}\right)$
multiplied by the proper isospin factor (for isoscalar this is one,
for isovector mesons it is $i\vec{\tau}$ etc.). The exact expression
on the right hand side depends on the matrix $\Gamma$, spin of the
meson and is usually given as a twist expansion over all possible
Lorentz structures which may be constructed from $\Gamma,q,\vec{r}$
and the polarization vector $\epsilon(q)$.

\begin{figure}[htb]
 \includegraphics[scale=0.7]{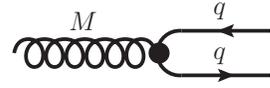} \caption{\label{fig:DA_Feynman}Diagram corresponding to the distribution amplitude~(\ref{eq:DA_definition})
in pQCD}

\end{figure}

In the leading order of $\alpha_{s}$ (which is justified for very
large $q^{2}$) the corresponding DA may be represented as a simple
diagram shown schematically in~Fig.~\ref{fig:DA_Feynman}. Then
we have

\begin{widetext}\begin{eqnarray}
\Phi_{M}\left(\beta,\vec{r};q\right) & \sim & \int d^{4}k\delta\left(\beta-k^{+}/q^{+}\right)e^{ik_{\perp}r_{\perp}}Tr\left[S(k)\Gamma_{M}S(k-q)\Gamma\right]=\label{eq:-3}\\
 & = & \int d^{2}k_{\perp}e^{ik_{\perp}r_{\perp}}\int dk^{-}\frac{f(k,q)}{\left(2k^{+}k^{-}-k_{\perp}^{2}-m^{2}+i0\right)\left(2\left(k^{+}-q^{+}\right)\left(k^{-}-q^{-}\right)-\left(k_{\perp}+q_{\perp}\right)^{2}-m^{2}+i0\right)},\nonumber \end{eqnarray}
 \end{widetext}where the function $f(k,q)=\left\langle \left(\hat{k}+m\right)\Gamma\left(\hat{k}+\hat{q}+m\right)\Gamma_{M}\right\rangle $
depends on the spins of mesons and Dirac matrices and is not important
for a moment.

\begin{figure}[htb]
 \includegraphics[scale=0.45]{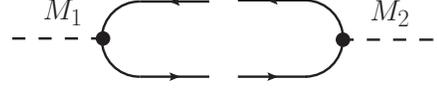}\caption{\label{fig:Overlap_Feynman}Diagram corresponding to overlap of the
distribution amplitudes}

\end{figure}

Taking the integral over $k^{-}$, we get \begin{eqnarray}
 &  & \int d^{4}k\delta\left(\beta-k^{+}/q^{+}\right)e^{ik_{\perp}r_{\perp}}Tr\left[S(k)\Gamma_{M}S(k-q)\Gamma\right]\label{eq:}\\
 & = & \theta\left(0\leq\beta\leq1\right)\int d^{2}k_{\perp}e^{ik_{\perp}r_{\perp}}\nonumber \\
 & \times & \frac{f(k,q)}{2\beta\left(1-\beta\right)q^{+}\left(2q^{+}q^{-}-\frac{k_{\perp}^{2}+m^{2}}{\beta}-\frac{\left(k_{\perp}+q_{\perp}\right)^{2}+m^{2}}{1-\beta}\right)}\nonumber \\
 & = & \theta\left(0\leq\beta\leq1\right)\int d^{2}k_{\perp}e^{ik_{\perp}r_{\perp}}\frac{f(k,q)}{2\left(\beta(1-\beta)q^{2}-\left(k_{\perp}^{2}+m^{2}\right)\right)}\nonumber \end{eqnarray}
 Straightforward evaluation of the overlap of the two functions $\left(\Phi_{A},\Phi_{\pi}\right)$
is quite tedious, however we may significantly simplify the evaluations
using completeness of the Dirac matrices, viz.\begin{equation}
\sum_{n}\Gamma_{\alpha\beta}^{(n)}\Gamma_{\alpha'\beta'}^{(n)}=\delta_{\alpha\alpha'}\delta_{\beta\beta'},\label{eq:-1}\end{equation}
 so the product of numerators of the two DAs is now converted to the
effective diagram shown in the Fig.(\ref{fig:Overlap_Feynman}).

Straightforward evaluation gives\begin{widetext}\begin{eqnarray}
F_{\mu}^{a\to\pi}\left(q,\Delta\right) & \sim & \sum_{\Gamma}\int d\beta d^{2}r\,\Phi_{\Gamma}^{(\pi)\dagger}\left(\beta,r_{\perp};q-\Delta\right)\Phi_{\Gamma}^{(a(\mu))}\left(\beta,r_{\perp};q\right)\sim\label{eq:-2}\\
 & \sim & 4mN_{c}\int d^{2}k_{\perp}\int d\beta\,\frac{q_{\mu}\left(4m^{2}+\Delta^{2}\right)-\Delta_{\mu}\left(4m^{2}+4k\cdot q+q^{2}\right)+2k_{\mu}\left(-2k\cdot q+\Delta^{2}-q^{2}+q\cdot\Delta\right)}{4\left(\beta(1-\beta)q^{2}-\left(k_{\perp}^{2}+m^{2}\right)\right)\left(\tilde{\beta}(1-\tilde{\beta})\left(q-\Delta\right)^{2}-\left(k_{\perp}^{2}+m^{2}\right)\right)}.\nonumber \end{eqnarray}
 \end{widetext}As we can see, the result is proportional to $\mathcal{O}(m)\sim\mathcal{O}\left(m_{\pi}^{2}\right)$
and thus is suppressed in the chiral limit. We expect that the same
result is valid for the nonperturbative DAs evaluated in the instanton
vacuum model.

\section{Numerical results}

\label{sec:Results}

The Bjorken variable $x=Q^{2}/\left(2\, p\cdot q\right)$, used at
high $Q^{2}$, is not appropriate at small $Q^{2}$, where it does
not have the meaning of a fractional quark momentum any more, and
may be very small even at low energies. For the case $Q^{2}=0$, where
the AR holds, $x$ defined in this way would be zero. Therefore, one
should rely on the phenomenological dipole cross section which depends
on energy, rather than $x$. At small $Q^{2}$ we employ the $s$-dependent
parametrization of the dipole cross section~\cite{kst2}, which is
similar to the $x$-dependent GBW parametrization \cite{GolecBiernat:1998js},
but is more suitable for soft processes \begin{eqnarray}
\sigma_{\bar{q}q}(r,s) & = & \sigma_{0}(s)\left(1-e^{-r^{2}/R_{0}^{2}(s)}\right),\label{eq:dipole-1}\\
R_{0}(s) & = & 0.88\, fm\times\left(\frac{s_{0}}{s}\right)^{0.14}.\label{eq:R0_standard}\end{eqnarray}
 These parameters and the scale $s_{0}=1000$~GeV$^{2}$ are fitted
to data on DIS, real photoproduction and $\pi p$ scattering. The
function $\sigma_{0}(s)$ is fixed by the condition \begin{equation}
\int d^{2}r\,\sigma_{\bar{q}q}(r,s)\int\limits _{0}^{1}d\beta\left|{\Psi}^{\pi}\left(\beta,\vec{r}\right)\right|^{2}\,=\,\sigma_{tot}^{\pi p}(s).\label{sig-tot}\end{equation}

\subsection{Corrections to the AR}

It was pointed out in \cite{kpss} that the AR (\ref{eq:Adler}) applied
to diffractive neutrino-production of pions, should be broken by absorptive
corrections, which affect the left- and right-hand sides of Eq.~(\ref{eq:Adler})
differently. Therefore, the AR Eq.~(\ref{eq:Adler}) cannot hold
universally, since the magnitude of absorptive corrections is target
dependent. The corrections reach maximum in the black-disc limit (e.g.
on heavy nuclei), where the AR is severely broken as was demonstrated
above in Sect.~\ref{sec:Overlap-details}. It was revealed in \cite{kpss}
that the AR is not accurate even on a proton target, a deviation of
about $30\%$ was estimated for diffractive pion production on a proton.
Unfortunately, the baseline for comparison of the left and right-hand
sides of Eq.~(\ref{eq:Adler}) is ill defined, it is not clear whether
the AR should hold with or without absorptive corrections.

The dipole phenomenology, which is adjusted to data, is free of this
uncertainty, it does not need to be corrected for absorption. One
can calculate the left and right-hand sides of Eq.~(\ref{eq:Adler})
on the same footing. Important is to use the same $\sigma_{tot}^{\pi p}(s)$
in (\ref{sig-tot}), as in the right-hand side of (\ref{eq:Adler}).
We employ only the pomeron part of the cross section parametrized
as $\sigma_{tot}^{\pi p}(s)=23.6mb\times(s/s_{0})^{0.08}$.

Calculation of the left and right-hand side of Eq.~(\ref{eq:Adler})
clearly demonstrates that absorptive corrections affect them differently.
Indeed, the amplitude in the left-hand side of (\ref{eq:Adler}) is
given by Eqs.~(\ref{eq:disgma_XSection}) and (\ref{eq:W_amplitude_defintiion}).
The first term in the dipole cross section (\ref{eq:dipole-1}) is
independent of $r$, therefore its contribution to neutrino-production
amplitudes vanishes, as was demonstrated in Sect.~\ref{sec:Overlap-details}.
So the production amplitude is suppressed by the second exponential
term in (\ref{eq:dipole-1}). This suppression becomes stronger with
energy, since $R(s)$ decreases, and at very high energies in the
Froissart regime the left-hand side of (\ref{eq:Adler}) vanishes.
At the same time, the pion-proton cross section Eq.~(\ref{sig-tot})
is dominated by the first term in (\ref{eq:dipole-1}) and reaches
maximum in the Froissart regime.

\begin{figure}[htb]
 \includegraphics[scale=0.4]{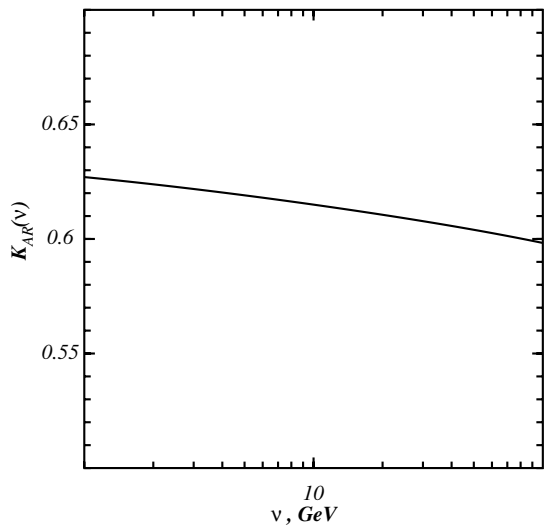}\caption{\label{fig:AxPi}Ratio of the cross-sections calculated within the
color dipole model and using the AR ~(\ref{eq:Adler}) at $\Delta_{\perp}=0$.}

\end{figure}

Now we are in a position to evaluate the accuracy of the AR for diffractive
neutrino-production of pions on protons. In Figure~\ref{fig:AxPi}
we plotted the ratio of the cross-sections calculated with the color
dipole model (left-hand side of (\ref{eq:Adler})) and using the AR
(right-hand side of (\ref{eq:Adler})), \begin{equation}
K_{AR}(s)=\left.\frac{d\sigma_{dipole}/dt\, d\nu\, dQ^{2}}{d\sigma_{AR}/dt\, d\nu\, dQ^{2}}\right|_{Q^{2}=0,\, t=0},\label{eq:R_PCAC}\end{equation}
 as was defined in \cite{kpss}.

As was expected, $K_{AR}<1$ due to different structures of the absorptive
corrections to diffractive pion production and elastic pion scattering
cross-sections. The deviation of $K_{AR}$ from unity is significant,
even somewhat larger than was estimated in \cite{kpss}. The ratio
is falling at high energies towards the Froissart limit, where it
eventually vanishes when $R_{0}(s)\to0$.

\subsection{Predicted cross sections}

Most of the data on neutrino-production of pions on protons have been
available so far only at energies close to the resonance region \cite{Kopeliovich:1992ym}.
Data at higher energies are scarce and have rather low statistics~\cite{Bell:1978qu,Allen:1985ti}.
Because the dipole formalism should not be trusted at low energies,
we provide predictions for the energy range of the ongoing experiment
Minerva at Fermilab~\cite{Drakoulakos:2004gn,McFarland:2006pz}. 

The $Q^{2}$ dependence of the diffractive cross section deserves
special attention. It would be very steep at small $Q^{2}$, if the
pion dominance were real. However, since the pion pole is terminated
due to conservation of the lepton current, the $Q^{2}$ dependence
is controlled by heavier singularities. In the approximation of an
effective singularity at $Q^{2}=-M^{2}$ \cite{kpss} one should expect
the dipole form $\propto(Q^{2}+M^{2})^{-2}$. Within the dispersion
approach the effective mass scale $M$ is expected to be of the order
of $1\,$GeV \cite{Belkov:1986hn,Kopeliovich:1992ym,kpss}. Within
the dipole description the $Q^{2}$ dependence is controlled by the
the IVM mass scale, which is of the order of $700$ MeV.

\begin{figure}[htb]
 \includegraphics[scale=0.4]{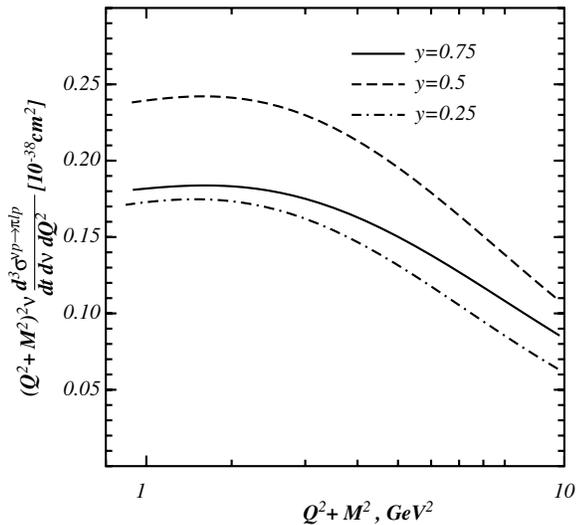}
\caption{\label{fig:Pion-proton-diff-q}The $Q^{2}$-dependence of the cross
section of diffractive neutrino-production of pions scaled by factor
$(Q^{2}+M^{2})^{2}$ at neutrino energy $E_{\nu}=20$~GeV and different $y$.
The mass parameter $M=0.91\,$GeV is adjusted to minimize the variations
of the scaled cross section at small $Q^{2}$.}

\end{figure}

In Fig.~\ref{fig:Pion-proton-diff-q}, we plot the forward diffractive
neutrino cross section scaled by the factor $(Q^{2}+M^{2})^{2}$,
where the parameter $M$ is adjusted in a way to provide a flat $Q^{2}$
dependence at $Q^{2}<2\,$GeV$^{2}$. %

Indeed, we found that at $M=0.91\,$GeV the scaled cross section is
constant up to rather large $Q^{2}\sim3\,$GeV$^{2}$, but substantially
deviates from the dipole form at larger $Q^{2}$.

\begin{figure}[htb]
 \includegraphics[scale=0.4]{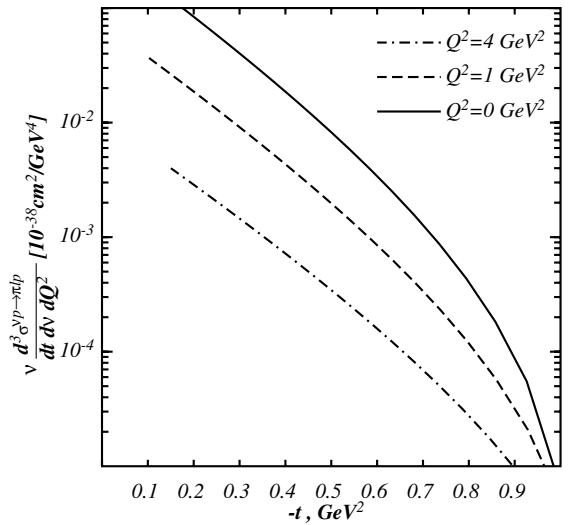} \caption{\label{fig:Pion-proton-diff-t}The $t$-dependence of the cross section
of diffractive neutrino-production of pions at different $Q^{2}$
for neutrino energy $E_{\nu}=20$~GeV and $y=0.5$.}

\end{figure}

The $t$-dependence of the cross section is controlled by the employed
model Eqs.~(\ref{eq:ImF}) for impact parameter dependence of the
dipole amplitude. The results for $t$-dependence of the invariant
cross-section are shown in Fig.~\ref{fig:Pion-proton-diff-t} for
several fixed values of $E_{\nu}$ and $Q^{2}=4\,$GeV$^{2}$. For
this calculation we fixed $y=0.5$. %

\begin{figure}[htb]
 \includegraphics[scale=0.4]{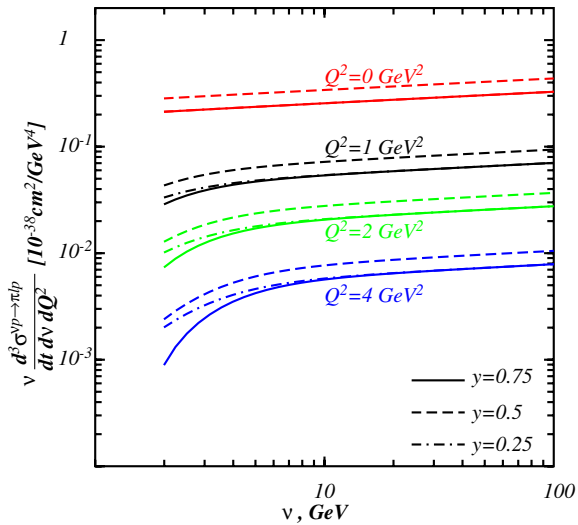}
\caption{\label{fig:Pion-proton-diff-nu}[Color online] Forward neutrino-production
cross-section of pions as function of $\nu$ at several fixed values
of $y$ and $Q^{2}$.}

\end{figure}

 The forward invariant cross-section Eq.~(\ref{eq:disgma_XSection})
of diffractive neutrino-production of pions on protons is depicted
in the Fig.~(\ref{fig:Pion-proton-diff-nu}) as function of $\nu$
at several fixed values of $y$ and $Q^{2}$.

These calculations performed in the dipole approach are controlled
in Eq.~(\ref{eq:W_amplitude_defintiion}) by the light-cone DAs of
the axial current and pion, which we extended to the soft interaction
regime basing on the instanton vacuum model. It worth reminding that
neither s-channel resonances, nor reggeons are included in the parametrization
(\ref{eq:dipole-1}) of the universal dipole cross section, so the
results are trustable only at sufficiently high energy $\nu$.

Experimental data for neutrino-production cross section are usually
presented as function of neutrino energy $E_{\nu}$ integrated over
$\nu$. Unfortunately, in this form one cannot separate physics of
low and high energies. Indeed, the integration over $\nu$ results
in the finite contribution of small-$\nu$ region, which is dominated
by $s$-channel resonances. This small-$\nu$ contribution  is constant
at any high neutrino energy $E_{\nu}$ and its magnitude is comparable
with the diffractive part.

Usually in low statistics experiments one integrates the multi-dimensional distributions
presenting the results as function of one variable. As such a variable we chose the
c.m. energy of the diffraction process, $W=\sqrt{m_N^2-Q^2+2m_N\nu}$.
Then we calculate the $W$-distribution as,
\begin{equation}
\frac{d\sigma}{dW}=2W\int d\nu\, dt\, dQ^{2}\delta\left(W^{2}-(p+q)^{2}\right)\frac{d\sigma}{d\nu\, dt\, dQ^{2}},\label{eq:dsdW}\end{equation}

\begin{figure*}[t]
\includegraphics[scale=0.4]{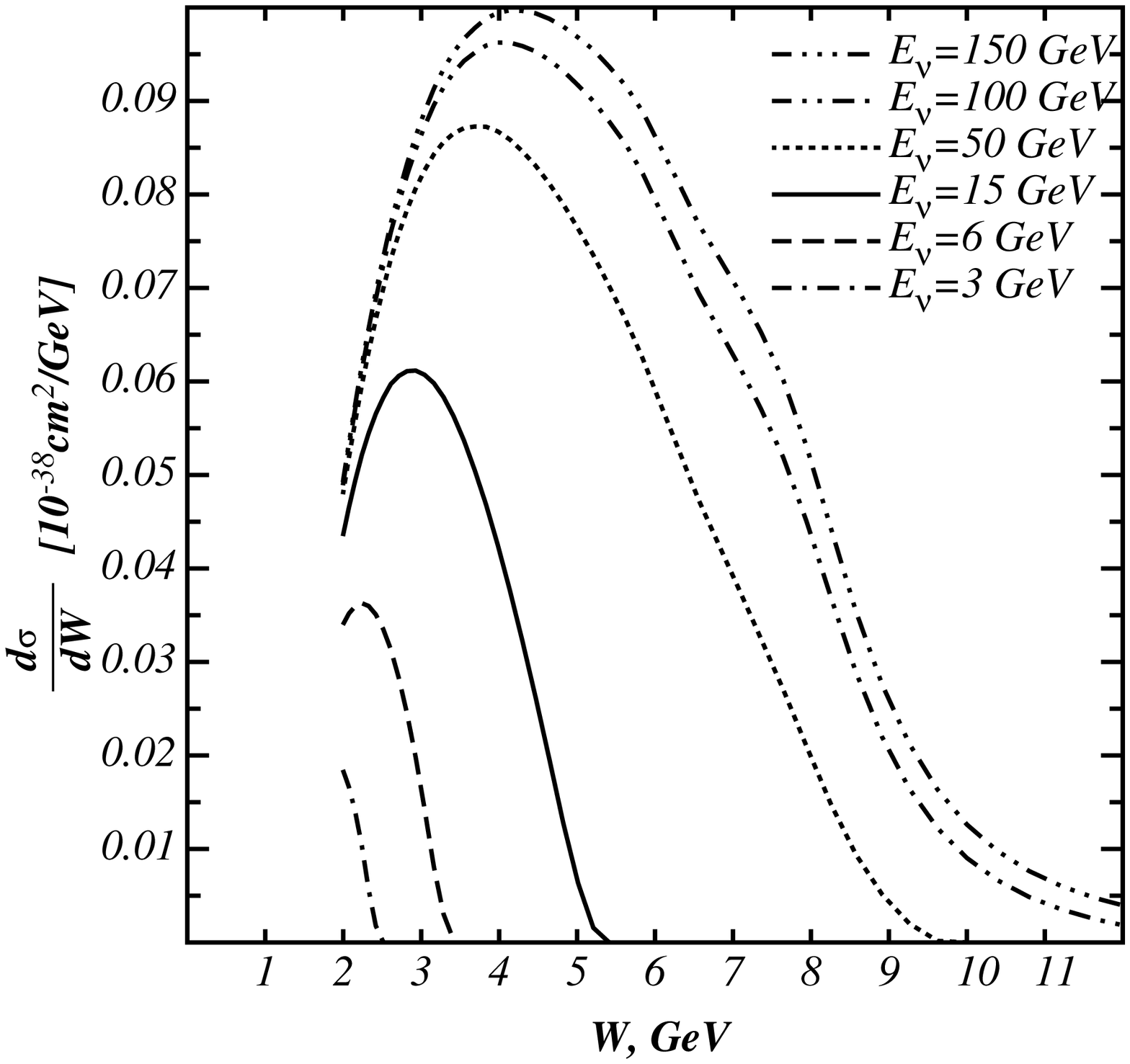}\qquad\includegraphics[scale=0.4]{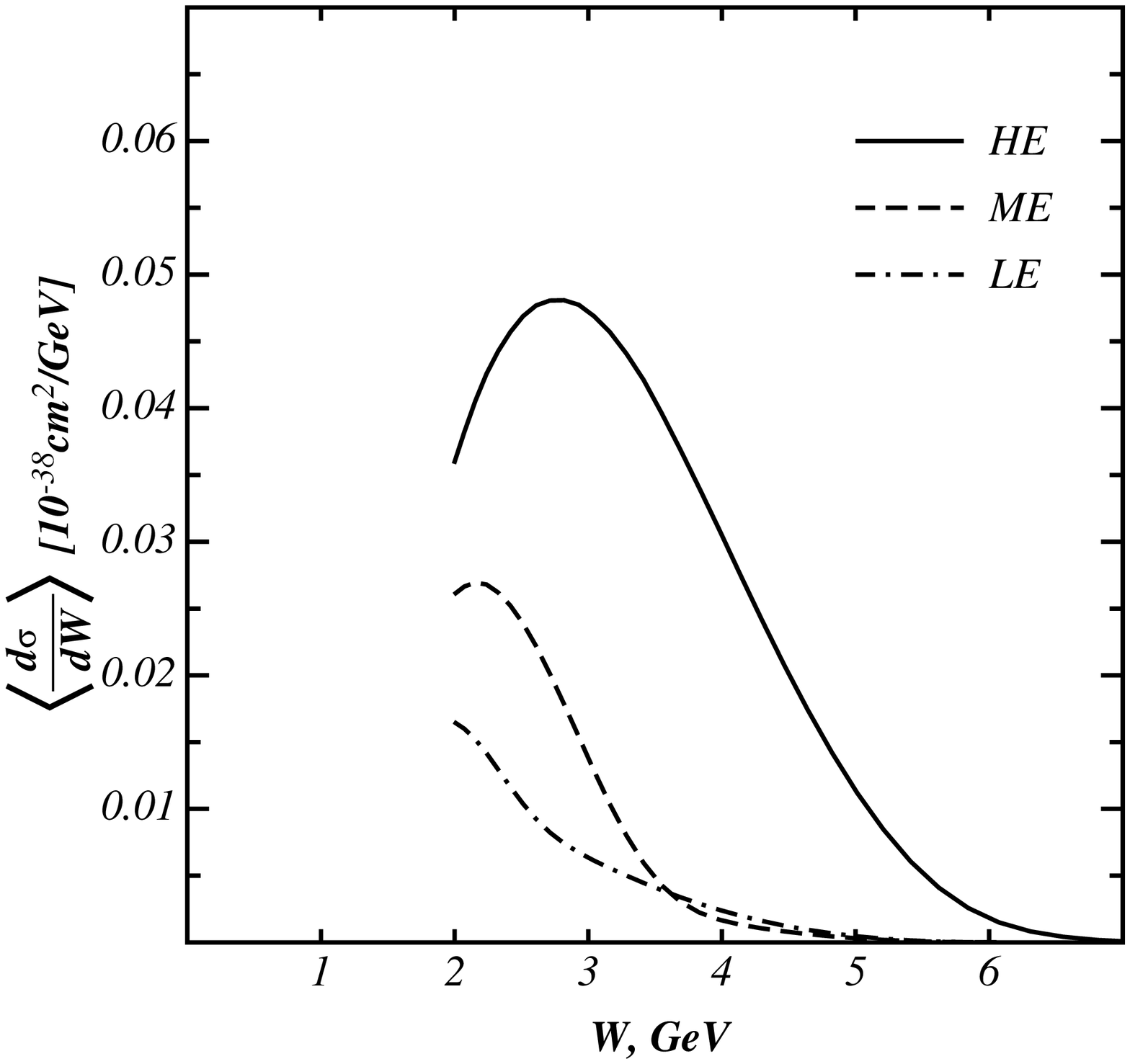}\caption{\label{fig:Neutrino-proton_tot} Left: Cross-section of diffractive
neutrino-production  $d\sigma/dW$ as function of $W$, for fixed
neutrino energies $E_{\nu}$. Right: the same cross-section $\langle d\sigma/dW\rangle$
weighted with the neutrino spectrum from Minerva~\cite{Drakoulakos:2004gn},
see~Eq.~(\ref{eq:Ave_dsdW}) for exact definition.}

\end{figure*}

In the left pane of the Figure~\ref{fig:Neutrino-proton_tot} we plotted
$W$-dependence of the cross-section~(\ref{eq:dsdW}) for several fixed
values of $E_{\nu}$. We see that the $W$-dependence significantly varies with $E_\nu$,
therefore one should average the cross section Eq.~(\ref{eq:dsdW}) weighted with a realistic neutrino energy distribution.
 \begin{equation}
\left\langle \frac{d\sigma}{dW}\right\rangle =\int dE_{\nu}\,\rho\left(E_{\nu}\right)\frac{d\sigma}{dW},\label{eq:Ave_dsdW}\end{equation}
 where the neutrino spectrum $\rho\left(E_{\nu}\right)$ is normalized
as \begin{equation}
\int dE_{\nu}\,\rho\left(E_{\nu}\right)=1.\label{eq:spec_norm}\end{equation}

As an example, we performed calculations with the neutrino energy spectrum of the MINERvA experiment~\cite{Drakoulakos:2004gn}.
We considered three different $E_\nu$-distributions corresponding to low
(LE), medium (ME) and high energy (HE) beam configurations.
The results are depicted in the right pane of the Figure~\ref{fig:Neutrino-proton_tot}.

\begin{figure}[t]
\includegraphics[scale=0.4]{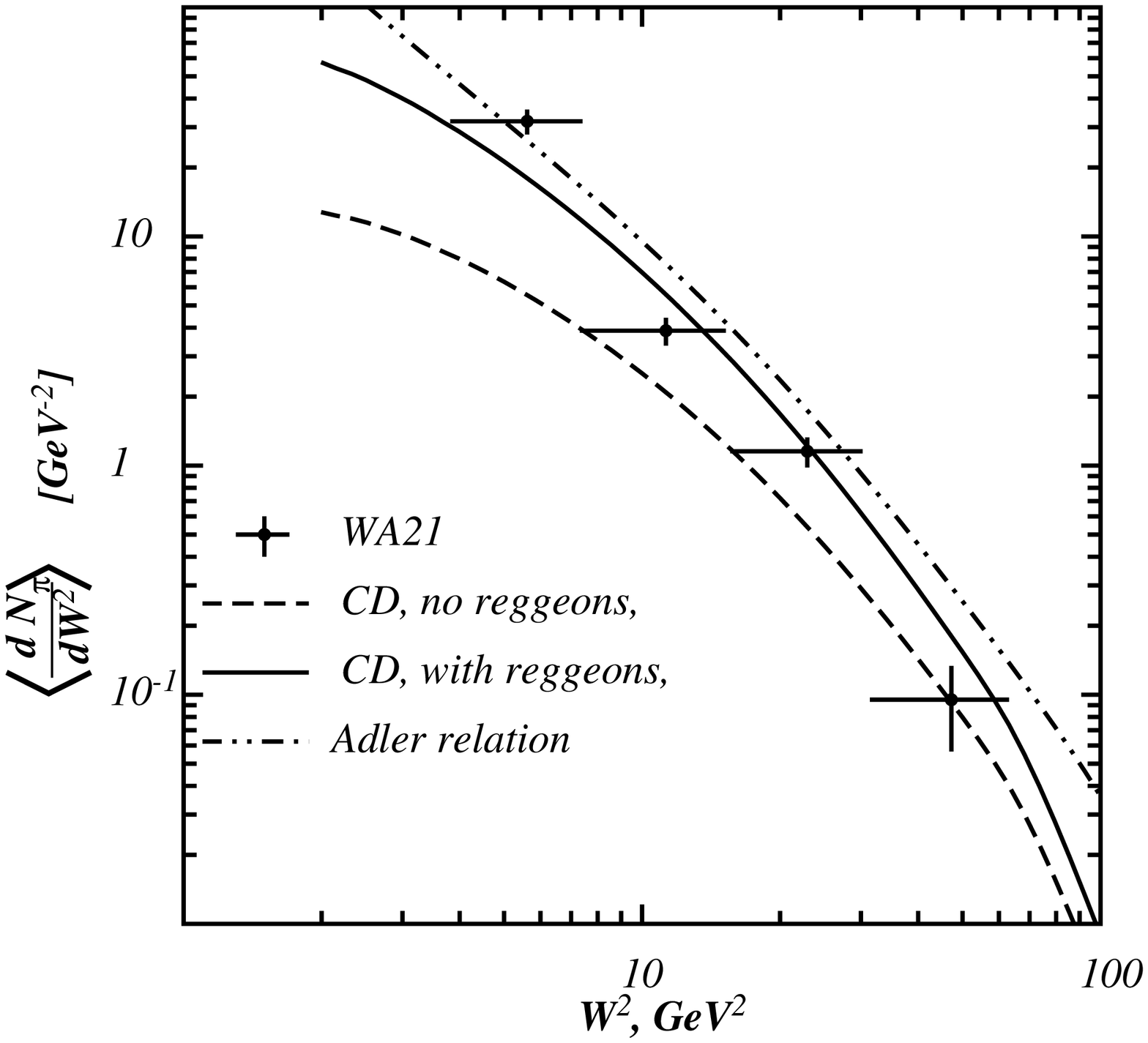}
\caption{\label{fig:WA21Comparison} Comparison of the color dipole prediction with experimental data from~\cite{Allen:1985ti}. ``CD'' stands for the color dipole model, either with or without reggeons, ``Adler relation'' stands for the evaluation using Adler relation~(\ref{eq:Adler}) extrapolated to nonzero $Q^2$ using dipole-type dependence for the $Q^2$-dependence, $\sim m_A^4/(m_A^2+Q^2)^2$.}
\end{figure}
We also compared our results for the $W$-distribution of neutrino
diffractive events with data of the WA21 experiment at CERN~\cite{Allen:1985ti}.
We performed averaging over neutrino energy with the spectrum $\rho(E_\nu)$
given in~\cite{Jones:1987wi}. The results are depicted by dashed curve in Fig.~\ref{fig:WA21Comparison}.
Since the low energy region is affected by reggeons, which we have
neglected so far, we added their contribution to the dipole cross section~(\ref{sig-tot}).

\begin{equation}
\sigma^{\pi p}_{tot}(s)=[13.6 s^{0.08}+19.2 s^{-0.45}] mb
\label{sigma_with_reggeons}
\end{equation}

The result shown by solid curve describes the data much better.
For comparison we plotted also the prediction based on the AR with the
realistic pion-proton cross section Eq.~(\ref{sigma_with_reggeons}).

\section{Summary}

\label{summary}

We developed the dipole description for high-energy neutrino interaction,
in particular at low $Q^{2}$, whether the PCAC hypothesis plays important
role. This approach is alternative to the conventional one based on
the dispersion relation for the $Q^{2}$ dependent amplitude axial
current interaction. While the latter faces the problem of lacking
experimental information on most of the diffractive diagonal and off-diagonal
amplitudes, the dipole formalism is free of these difficulties. Besides,
one can employ the universal dipole cross section (see Eq.~(\ref{eq:W_amplitude_defintiion})),
well fixed by numerous data for interactions of the vector current
in electromagnetic processes (DIS, photoproduction, etc.).

The important challenge of the dipole description is the construction
of the current distribution amplitudes at small $Q^{2}$, where the
nonperturbative effects are unavoidable. We calculated the distribution
amplitudes for the axial current (Sect.~\ref{axialDA}) and for the
pion (Sect.~\ref{pion}) on the same footing, within the instanton
vacuum model (Sect.~\ref{ivm}). The model possesses the chiral symmetry
properties, what guarantees a correct, controlled by PCAC behavior
at small $Q^{2}$.

Although the dipole approach does not involve explicitly the intermediate
hadronic states, absence of the pion pole can be tested in the "black-disc"
regime, where all the partial elastic amplitude saturate at the unitarity
bound. Indeed, the direct calculation performed in Sect.~\ref{sec:Overlap-details}
confirmed that diffractive pion production ceases, what may happen
only if the pion pole does not contribute.

The dipole description also offers an unbiased way to test the AR
on a proton target. This relation is expected to be broken by absorptive
corrections \cite{kpss}, which are implicitly included in the phenomenological
dipole cross section. We rely on the dipole cross section parametrized
in the saturation form Eq.~(\ref{eq:dipole-1}), well confirmed by
data for electromagnetic processes \cite{GolecBiernat:1998js}. We
found a significant, about $40\%$ deviation from the AR on a proton
target (see Fig.~\ref{fig:AxPi}).

A much stronger breakdown of the AR is expected for nuclei \cite{kpss},
and we plan to evaluate those effects employing the techniques developed
here. \\

\begin{acknowledgments}
We are grateful to Genya Levin for numerous discussions and improving
comments, and to Sasha Dorokhov for the fruitful discussion of the
pion distribution amplitudes. We thank Jorge Morfin for the interest
to our results and an advice regarding comparison with data. This
work was supported in part by Fondecyt (Chile) grants No. 1090073,
1090291 and 1100287, and by Conicyt-DFG grant No. 084-2009. 
\end{acknowledgments}
\appendix

\section{Distribution amplitudes in the instanton vacuum model}

\label{sec:DADetails}

In this section we present the results of calculation of the DAs for
the axial current and pion performed in the instanton vacuum model
(IVM).

\subsection{Axial DAs}

\label{sec:axialDA}

There are eight independent axial DAs defined in~(\ref{eq:AWF-mu5}-\ref{eq:AWF-5}),

\begin{widetext}\begin{eqnarray}
\left\langle 0\left|\bar{\psi}\left(y\right)\gamma_{\mu}\gamma_{5}\psi\left(x\right)\right|A(q)\right\rangle  & = & if_{A}^{2}\int_{0}^{1}du\, e^{i(up\cdot y+\bar{u}p\cdot x)}\nonumber \\
 & \times & \left[p_{\mu}\frac{e^{(\lambda)}\cdot z}{p\cdot z}\Phi_{||}(u)+e_{\mu}^{(\lambda=\perp)}g_{\perp}^{(a)}(u)-\frac{1}{2}z_{\mu}\frac{e^{(\lambda)}\cdot z}{(p\cdot z)^{2}}f_{A}^{2}g_{3}(u)\right],\label{eq:AWF-mu5-a}\end{eqnarray}

\begin{eqnarray}
\left\langle 0\left|\bar{\psi}\left(y\right)\gamma_{\mu}\psi\left(x\right)\right|A(q)\right\rangle  & = & -if_{A}^{2}\epsilon_{\mu\nu\rho\sigma}e_{\nu}^{(\lambda)}p_{\rho}z_{\sigma}\int_{0}^{1}du\, e^{i(up\cdot y+\bar{u}p\cdot x)}\frac{g_{\perp}^{(v)}(u)}{4}\label{eq:AWF-mu-a}\end{eqnarray}

\begin{eqnarray}
\left\langle 0\left|\bar{\psi}\left(y\right)\sigma_{\mu\nu}\gamma_{5}\psi\left(x\right)\right|A(q)\right\rangle  & = & f_{A}\int_{0}^{1}du\, e^{i(up\cdot y+\bar{u}p\cdot x)}\biggl[\left(e_{\mu}^{(\lambda=\perp)}p_{\nu}-e_{\nu}^{(\lambda=\perp)}p_{\mu}\right)\Phi_{\perp}(u)\nonumber \\
 & + & \left.\frac{e^{(\lambda)}\cdot z}{(p\cdot z)^{2}}f_{A}^{2}\left(p_{\mu}z_{\nu}-p_{\nu}z_{\mu}\right)h_{||}^{(t)}(u)+\frac{1}{2}\left(e_{\mu}^{(\lambda)}z_{\nu}-e_{\nu}^{(\lambda)}z_{\mu}\right)\frac{f_{A}^{2}}{p\cdot z}h_{3}(u)\right],\label{eq:AWF-munu-a}\end{eqnarray}

\begin{eqnarray}
\left\langle 0\left|\bar{\psi}\left(y\right)\gamma_{5}\psi\left(x\right)\right|A(q)\right\rangle  & = & f_{A}^{3}e^{(\lambda)}\cdot z\int_{0}^{1}du\, e^{i(up\cdot y+\bar{u}p\cdot x)}\frac{h_{||}^{(p)}(u)}{2},\label{eq:AWF-5-a}\end{eqnarray}

After tedious but straightforward calculations we arrive at,

{\small \begin{eqnarray}
 &  & \Phi_{||}\left(u,\vec{r}_{\perp}\right)=\frac{1}{if_{A}^{2}n\cdot e^{(\lambda)}}\int\frac{dz}{2\pi}e^{i(u-1/2)z}\left\langle 0\left|\bar{\psi}\left(-\frac{z}{2}n-\frac{\vec{r}_{\perp}}{2}\right)\hat{n}\gamma_{5}\psi\left(\frac{z}{2}n+\frac{\vec{r}_{\perp}}{2}\right)\right|A^{(\lambda)}(q)\right\rangle =\label{eq:Phi_parallel_0}\\
 & = & \frac{8N_{c}}{if_{A}^{2}}\int\frac{dl^{-}d^{2}l_{\perp}}{(2\pi)^{4}}e^{-i\vec{l}_{\perp}\cdot\vec{r}_{\perp}}\times\\
 & \times & \left[\frac{\mu(l)\mu(l+q)+l_{\perp}^{2}+\left(\frac{1}{4}-u^{2}\right)q^{2}}{\left(l^{2}+\mu^{2}(l)\right)\left(\left(l+q\right)^{2}+\mu^{2}(l+q)\right)}+\frac{M\left(f(l+q)-f(l)\right)^{2}\left(2l^{-}-q^{2}\left(u-\frac{1}{2}\right)\right)\left(\mu(l)\left(u+\frac{1}{2}\right)-\mu(l+q)\left(u-\frac{1}{2}\right)\right)}{\left((l+q)^{2}-l^{2}\right)\left(l^{2}+\mu^{2}(l)\right)\left(\left(l+q\right)^{2}+\mu^{2}(l+q)\right)}\right]_{l^{+}=\left(u-\frac{1}{2}\right)q^{+}}\end{eqnarray}
 }{\small \par}

{\small \begin{eqnarray*}
\Phi_{\perp}\left(u,\vec{r}_{\perp}\right) & = & \frac{1}{2f_{A}}\int\frac{dz}{2\pi}e^{i(u-1/2)z}\left\langle 0\left|\bar{\psi}\left(-\frac{z}{2}n\right)\sigma_{\nu\rho}\gamma_{5}\left(e_{\nu}^{(\lambda=\perp)}n_{\rho}-e_{\rho}^{(\lambda=\perp))}n_{\nu}\right)\psi\left(\frac{z}{2}n\right)\right|A^{(\lambda=\perp)}(q)\right\rangle =\\
 & = & \frac{4N_{c}}{\pi f_{A}}\int\frac{dl^{-}d^{2}l_{\perp}}{(2\pi)^{4}}e^{-i\vec{l}_{\perp}\cdot\vec{r}_{\perp}}\times\\
 & \times & \left[\frac{\left(u-\frac{1}{2}\right)\mu(l+q)+\left(u+\frac{1}{2}\right)\mu(l)}{\left(l^{2}+\mu^{2}(l)\right)\left(\left(l+q\right)^{2}+\mu^{2}(l+q)\right)}-\frac{l_{\mu_{\perp}}^{2}}{(l+q)^{2}-l^{2}}\frac{M\left(f(l+q)-f(l)\right)^{2}}{\left(l^{2}+\mu^{2}(l)\right)\left(\left(l+q\right)^{2}+\mu^{2}(l+q)\right)}\right]_{l^{+}=\left(u-\frac{1}{2}\right)q^{+}}\end{eqnarray*}
 }{\small \par}

{\small \begin{eqnarray*}
g_{\perp}^{(a)}\left(u,\vec{r}_{\perp}\right) & = & \frac{1}{if_{A}^{2}}\int\frac{dz}{2\pi}e^{i(u-1/2)z}\left\langle 0\left|\bar{\psi}\left(-\frac{z}{2}n\right)\hat{e}^{(\lambda=\perp)}\gamma_{5}\psi\left(\frac{z}{2}n\right)\right|A^{(\lambda=\perp)}(q)\right\rangle =\\
 & = & \frac{4N_{c}}{i\pi f_{A}^{2}}\int\frac{dl^{-}d^{2}l_{\perp}}{(2\pi)^{4}}e^{-i\vec{l}_{\perp}\cdot\vec{r}_{\perp}}\times\\
 & \times & \left[\frac{\left(\mu(l)\mu(l+q)-l^{2}-l\cdot q\right)+l_{\perp}^{2}}{\left(l^{2}+\mu^{2}(l)\right)\left(\left(l+q\right)^{2}+\mu^{2}(l+q)\right)}-\frac{l_{\mu_{\perp}}^{2}}{(l+q)^{2}-l^{2}}\frac{M\left(f(l+q)-f(l)\right)^{2}\left(\mu(l+q)-\mu(l)\right)}{\left(l^{2}+\mu^{2}(l)\right)\left(\left(l+q\right)^{2}+\mu^{2}(l+q)\right)}\right]_{l^{+}=\left(u-\frac{1}{2}\right)q^{+}}\end{eqnarray*}
 }{\small \par}

{\small \begin{eqnarray*}
g_{\perp}^{(v)}\left(u,\vec{r}_{\perp}\right) & = & \frac{4i}{f_{A}^{2}}Coefficient\left(\int\frac{dz}{2\pi}e^{i(u-1/2)z}\left\langle 0\left|\bar{\psi}\left(-\frac{z}{2}n\right)\hat{e}^{(\lambda=\perp)}\gamma_{\mu}\psi\left(\frac{z}{2}n\right)\right|A^{(\lambda=\perp)}(q)\right\rangle ,\epsilon_{\mu\nu\rho\sigma}e_{\nu}^{(\lambda)}p_{\rho}n_{\sigma}\right)=\\
 & = & \frac{32N_{c}}{f_{A}^{2}}\int\frac{dl^{+}}{2\pi}\frac{1}{q^{+}}\delta\left(\frac{l^{+}}{q^{+}}-u+\frac{1}{2}\right)\int\frac{dl^{-}d^{2}l_{\perp}}{(2\pi)^{4}}e^{-i\vec{l}_{\perp}\cdot\vec{r}_{\perp}}\times\\
 & \times & \left[\frac{q\cdot l-q^{2}\left(u-\frac{1}{2}\right)}{\left(l^{2}+\mu^{2}(l)\right)\left(\left(l+q\right)^{2}+\mu^{2}(l+q)\right)}\right]_{l^{+}=\left(u-\frac{1}{2}\right)q^{+}}\end{eqnarray*}
 }{\small \par}

{\small \begin{eqnarray*}
h_{||}^{(t)}\left(u,\vec{r}_{\perp}\right) & = & -\frac{1}{2f_{A}^{3}e^{(\lambda)}\cdot n}\int\frac{dz}{2\pi}e^{i(u-1/2)z}\left\langle 0\left|\bar{\psi}\left(-\frac{z}{2}n\right)\sigma_{\nu\rho}\gamma_{5}\left(p_{\nu}n_{\rho}-p_{\rho}n_{\nu}\right)\psi\left(\frac{z}{2}n\right)\right|A^{(\lambda)}(q)\right\rangle =\\
 & = & \frac{8N_{c}}{f_{A}^{3}}\int\frac{dl^{-}d^{2}l_{\perp}}{(2\pi)^{4}}e^{-i\vec{l}_{\perp}\cdot\vec{r}_{\perp}}\times\\
 & \times & \left[\frac{\mu(l+q)\left(l^{-}+\left(u-\frac{1}{2}\right)\frac{q^{2}}{2}\right)+\mu(l)\left(l^{-}+\left(u+\frac{3}{2}\right)\frac{q^{2}}{2}\right)}{\left(l^{2}+\mu^{2}(l)\right)\left(\left(l+q\right)^{2}+\mu^{2}(l+q)\right)}+\frac{2\left(\left(l^{-}\right)^{2}-\left(u-\frac{1}{2}\right)^{2}\frac{q^{4}}{4}\right)M\left(f(l+q)-f(l)\right)^{2}}{\left((l+q)^{2}-l^{2}\right)\left(l^{2}+\mu^{2}(l)\right)\left(\left(l+q\right)^{2}+\mu^{2}(l+q)\right)}\right]_{l^{+}=\left(u-\frac{1}{2}\right)q^{+}}\end{eqnarray*}
 }{\small \par}

{\small \begin{eqnarray*}
h_{||}^{(p)}\left(u,\vec{r}_{\perp}\right) & = & \frac{1}{\left(f_{A}^{3}e^{(\lambda)}\cdot n\right)}\int\frac{dz}{2\pi}e^{i(u-1/2)z}\left\langle 0\left|\bar{\psi}\left(-\frac{z}{2}n\right)\hat{e}^{(\lambda=\perp)}\gamma_{5}\psi\left(\frac{z}{2}n\right)\right|A^{(\lambda=\perp)}(q)\right\rangle =\\
 & = & -\frac{8N_{c}}{f_{A}^{3}}\int\frac{dl^{+}}{2\pi}\frac{1}{q^{+}}\delta\left(\frac{l^{+}}{q^{+}}-u+\frac{1}{2}\right)\int\frac{dl^{-}d^{2}l_{\perp}}{(2\pi)^{4}}e^{-i\vec{l}_{\perp}\cdot\vec{r}_{\perp}}\\
 & \times & \left[\frac{\left(\mu(l)-\mu(l+q)\right)\left(l^{-}+\left(\frac{1}{2}-u\right)\frac{q^{2}}{2}\right)}{\left(l^{2}+\mu^{2}(l)\right)\left(\left(l+q\right)^{2}+\mu^{2}(l+q)\right)}\right.\\
 & - & \left.\frac{\left(-l_{\perp}^{2}+2ul^{-}+\left(u-\frac{1}{2}\right)\frac{q^{2}}{2}+\mu(l)\mu(l+q)\right)}{(l+q)^{2}-l^{2}}\frac{M\left(f(l+q)-f(l)\right)^{2}\left(2l^{-}-q^{2}\left(u-\frac{1}{2}\right)\right)}{\left(l^{2}+\mu^{2}(l)\right)\left(\left(l+q\right)^{2}+\mu^{2}(l+q)\right)}\right]_{l^{+}=\left(u-\frac{1}{2}\right)q^{+}}\end{eqnarray*}
 }{\small \par}

{\small \begin{eqnarray*}
 &  & g_{3}\left(u,\vec{r}_{\perp}\right)=-\frac{2}{f_{A}^{2}e^{(\lambda)}\cdot n}\int\frac{dz}{2\pi}e^{i(u-1/2)z}\left\langle 0\left|\bar{\psi}\left(-\frac{z}{2}n\right)\hat{p}\gamma_{5}\psi\left(\frac{z}{2}n\right)\right|A^{(\lambda)}(q)\right\rangle =\\
 & = & -\frac{8N_{c}}{f_{A}^{2}}\int\frac{dl^{-}d^{2}l_{\perp}}{(2\pi)^{4}}e^{-i\vec{l}_{\perp}\cdot\vec{r}_{\perp}}\times\\
 & \times & \left[\frac{2(l^{-})^{2}+l^{-}q^{2}-\left(\mu(l)\mu(l+q)+l_{\perp}^{2}\right)\frac{q^{2}}{2}}{\left(l^{2}+\mu^{2}(l)\right)\left(\left(l+q\right)^{2}+\mu^{2}(l+q)\right)}-\frac{\left(\mu(l)\left(l^{-}+\frac{q^{2}}{2}\right)-\mu(l+q)l^{-}\right)}{\left((l+q)^{2}-l^{2}\right)}\frac{M\left(f(l+q)-f(l)\right)^{2}\left(l^{-}-\frac{q^{2}}{2}\left(u-\frac{1}{2}\right)\right)}{\left(l^{2}+\mu^{2}(l)\right)\left(\left(l+q\right)^{2}+\mu^{2}(l+q)\right)}\right]_{l^{+}=\left(u-\frac{1}{2}\right)q^{+}}\end{eqnarray*}
 }{\small \par}

{\small \begin{eqnarray*}
h_{3}\left(u,\vec{r}_{\perp}\right) & = & \frac{1}{f_{A}^{3}}\int\frac{dz}{2\pi}e^{i(u-1/2)z}\left\langle 0\left|\bar{\psi}\left(-\frac{z}{2}n\right)\sigma_{\nu\rho}\gamma_{5}\left(e_{\nu}^{(\lambda=\perp)}p_{\rho}-e_{\rho}^{(\lambda=\perp)}p_{\nu}\right)\psi\left(\frac{z}{2}n\right)\right|A^{(\lambda=\perp)}(q)\right\rangle =\\
 & = & \frac{16N_{c}}{f_{A}^{3}}\int\frac{dl^{-}d^{2}l_{\perp}}{(2\pi)^{4}}e^{-i\vec{l}_{\perp}\cdot\vec{r}_{\perp}}\times\\
 & \times & \left[\frac{l_{-}\mu(l+q)+\left(l_{-}+q_{-}\right)\mu(l)}{\left(l^{2}+\mu^{2}(l)\right)\left(\left(l+q\right)^{2}+\mu^{2}(l+q)\right)}-l_{\mu_{\perp}}^{2}\frac{M\left(f(l+q)-f(l)\right)^{2}}{\left(l^{2}+\mu^{2}(l)\right)\left(\left(l+q\right)^{2}+\mu^{2}(l+q)\right)}\right]_{l^{+}=\left(u-\frac{1}{2}\right)q^{+}}\end{eqnarray*}
 }\end{widetext}

\subsection{Pion DAs}

\label{pionDA}

For the pion there are four independent pion DAs defined in~(\ref{eq:piWF-mu5}-\ref{eq:piWF-munu}),\begin{eqnarray}
 &  & \left\langle 0\left|\bar{\psi}\left(y\right)\gamma_{\mu}\gamma_{5}\psi\left(x\right)\right|\pi(q)\right\rangle =if_{\pi}\int_{0}^{1}du\, e^{i(up\cdot y+\bar{u}p\cdot x)}\nonumber \\
 & \times & \left(p_{\mu}\phi_{2;\pi}(u)+\frac{1}{2}\frac{z_{\mu}}{(p\cdot z)}\psi_{4;\pi}(u)\right);\label{eq:piWF-mu5-a}\end{eqnarray}

\begin{eqnarray}
 &  & \left\langle 0\left|\bar{\psi}\left(y\right)\gamma_{5}\psi\left(x\right)\right|\pi(q)\right\rangle =-if_{\pi}\frac{m_{\pi}^{2}}{m_{u}+m_{d}}\nonumber \\
 & \times & \int_{0}^{1}du\, e^{i(up\cdot y+\bar{u}p\cdot x)}\phi_{3;\pi}^{(p)}(u);\label{eq:piWF-5-a}\end{eqnarray}

\begin{eqnarray}
 &  & \left\langle 0\left|\bar{\psi}\left(y\right)\sigma_{\mu\nu}\gamma_{5}\psi\left(x\right)\right|\pi(q)\right\rangle =-\frac{i}{3}f_{\pi}\frac{m_{\pi}^{2}}{m_{u}+m_{d}}\nonumber \\
 & \times & \int_{0}^{1}du\,\phi_{3;\pi}^{(\sigma)}(u)\,\frac{e^{i(up\cdot y+\bar{u}p\cdot x)}}{p\cdot z}\left(p_{\mu}z_{\nu}-p_{\nu}z_{\mu}\right).\label{eq:piWF-munu-a}\end{eqnarray}
 Eventually we arrive at the following structures in the pion DA,
\begin{widetext}

{\small \begin{eqnarray}
\phi_{2;\pi}\left(u,\vec{r}_{\perp}\right) & = & \frac{1}{if_{\pi}}\int\frac{dz}{2\pi}e^{i(u-1/2)z}\left\langle 0\left|\bar{\psi}\left(-\frac{z}{2}n-\frac{\vec{r}_{\perp}}{2}\right)\hat{n}\gamma_{5}\psi\left(\frac{z}{2}n+\frac{\vec{r}_{\perp}}{2}\right)\right|\pi(q)\right\rangle =\label{eq:phi_2;pi_0}\\
 & = & \frac{8N_{c}}{f_{\pi}}\int\frac{dl^{-}d^{2}l_{\perp}}{(2\pi)^{4}}e^{-i\vec{l}_{\perp}\cdot\vec{r}_{\perp}}\left[Mf(l)f(l+q)\frac{\mu(l)\left(u+\frac{1}{2}\right)-\mu(l+q)\left(u-\frac{1}{2}\right)}{\left(l^{2}+\mu^{2}(l)\right)\left(\left(l+q\right)^{2}+\mu^{2}(l+q)\right)}\right]_{l^{+}=\left(u-\frac{1}{2}\right)q^{+}};\nonumber \end{eqnarray}
 }{\small \par}

{\small \begin{eqnarray}
\psi_{4;\pi}\left(u,\vec{r}_{\perp}\right) & = & \frac{2}{if_{\pi}}\int\frac{dz}{2\pi}e^{i(u-1/2)z}\left\langle 0\left|\bar{\psi}\left(-\frac{z}{2}n-\frac{\vec{r}_{\perp}}{2}\right)\hat{p}\gamma_{5}\psi\left(\frac{z}{2}n+\frac{\vec{r}_{\perp}}{2}\right)\right|\pi(q)\right\rangle =\label{eq:psi_4;pi_0}\\
 & = & \frac{16N_{c}}{f_{\pi}}\int\frac{dl^{-}d^{2}l_{\perp}}{(2\pi)^{4}}e^{-i\vec{l}_{\perp}\cdot\vec{r}_{\perp}}\left[Mf(l)f(l+q)\frac{\mu(l)\left(l_{-}+q_{-}\right)-\mu(l+q)l_{-}}{\left(l^{2}+\mu^{2}(l)\right)\left(\left(l+q\right)^{2}+\mu^{2}(l+q)\right)}\right]_{l^{+}=\left(u-\frac{1}{2}\right)q^{+}};\nonumber \end{eqnarray}
 }{\small \par}

{\small \begin{eqnarray}
\phi_{3;\pi}^{(p)}\left(u,\vec{r}_{\perp}\right) & = & \frac{1}{f_{\pi}}\frac{m_{u}+m_{d}}{m_{\pi}^{2}}\int\frac{dz}{2\pi}e^{i(u-1/2)z}\left\langle 0\left|\bar{\psi}\left(-\frac{z}{2}n-\frac{\vec{r}_{\perp}}{2}\right)\gamma_{5}\psi\left(\frac{z}{2}n+\frac{\vec{r}_{\perp}}{2}\right)\right|\pi(q)\right\rangle =\label{eq:phi_3;pi_p_0}\\
 & = & \frac{8N_{c}}{f_{\pi}}\frac{m_{u}+m_{d}}{m_{\pi}^{2}}\int\frac{dl^{-}d^{2}l_{\perp}}{(2\pi)^{4}}e^{-i\vec{l}_{\perp}\cdot\vec{r}_{\perp}}\left[Mf(l)f(l+q)\frac{\mu(l)\mu(l+q)+l^{2}+l\cdot q}{\left(l^{2}+\mu^{2}(l)\right)\left(\left(l+q\right)^{2}+\mu^{2}(l+q)\right)}\right]_{l^{+}=\left(u-\frac{1}{2}\right)q^{+}};\nonumber \end{eqnarray}
 }{\small \par}

{\small \begin{eqnarray}
\phi_{3;\pi}^{(\sigma)}\left(u,\vec{r}_{\perp}\right) & = & \frac{3i}{2f_{\pi}}\frac{m_{u}+m_{d}}{m_{\pi}^{2}}\int\frac{dz}{2\pi}e^{i(u-1/2)z}\left\langle 0\left|\bar{\psi}\left(-\frac{z}{2}n-\frac{\vec{r}_{\perp}}{2}\right)\left(p_{\mu}n_{\nu}-p_{\nu}n_{\mu}\right)\sigma_{\mu\nu}\gamma_{5}\psi\left(\frac{z}{2}n+\frac{\vec{r}_{\perp}}{2}\right)\right|\pi(q)\right\rangle =\label{eq:phi_3;pi_sigma_0}\\
 & = & -\frac{24N_{c}}{f_{\pi}}\frac{m_{u}+m_{d}}{m_{\pi}^{2}}\int\frac{dl^{-}d^{2}l_{\perp}}{(2\pi)^{4}}e^{-i\vec{l}_{\perp}\cdot\vec{r}_{\perp}}\left[Mf(l)f(l+q)\frac{q_{+}l_{-}-\frac{q^{2}}{2}\left(u-\frac{1}{2}\right)}{\left(l^{2}+\mu^{2}(l)\right)\left(\left(l+q\right)^{2}+\mu^{2}(l+q)\right)}\right]_{l^{+}=\left(u-\frac{1}{2}\right)q^{+}}.\nonumber \end{eqnarray}
 }{\small \par}

\end{widetext} The details of calculation of the distribution amplitudes
will be presented elsewhere.

\subsection{Vector current DAs}

\label{vectorDA}

The vector current DAs were derived in~\cite{Dorokhov:2006qm},\begin{widetext}

\begin{eqnarray}
\left\langle 0\left|\bar{\psi}\left(y\right)\gamma_{\mu}\gamma_{5}\psi\left(x\right)\right|V(q)\right\rangle  & = & e_{q}f_{3\gamma}f_{\gamma}^{a}(q)\epsilon_{\mu\nu\rho\sigma}e_{\nu}^{(\lambda)}p_{\rho}z_{\sigma}\int_{0}^{1}du\, e^{i(up\cdot y+\bar{u}p\cdot x)}\psi_{\gamma}^{(a)}\left(u,q^{2}\right)\label{eq:VWF-mu5}\end{eqnarray}

\begin{eqnarray}
\left\langle 0\left|\bar{\psi}\left(y\right)\gamma_{\mu}\psi\left(x\right)\right|V(q)\right\rangle  & = & e_{q}f_{3\gamma}f_{\perp\gamma}^{v}(q)\int_{0}^{1}du\, e^{i(up\cdot y+\bar{u}p\cdot x)}\nonumber \\
 & \times & \left[p_{\mu}\left(e^{(\lambda)}\cdot n\right)\frac{f_{||\gamma}^{(v)}(q)}{f_{\perp\gamma}^{(v)}(q)}\phi_{||}\left(u,q^{2}\right)+e_{\mu}^{(\lambda=\perp)}\psi_{\perp\gamma}^{(v)}\left(u,q^{2}\right)+n_{\mu}\left(e^{(\lambda)}\cdot n\right)h_{\gamma}^{(v)}\left(u,q^{2}\right)\right]\end{eqnarray}

\begin{eqnarray}
\left\langle 0\left|\bar{\psi}\left(y\right)\sigma_{\mu\nu}\psi\left(x\right)\right|V(q)\right\rangle  & = & ie_{q}\langle\bar{q}q\rangle f_{\perp\gamma}^{t}\left(q^{2}\right)\int_{0}^{1}du\, e^{i(up\cdot y+\bar{u}p\cdot x)}\left[\left(e_{\mu}^{(\lambda=\perp)}p_{\nu}-e_{\nu}^{(\lambda=\perp)}p_{\mu}\right)\chi_{m}\phi_{\perp\gamma}\left(u,q^{2}\right)\right.\nonumber \\
 & + & \left.\left(e^{(\lambda)}\cdot n\right)\left(p_{\mu}n_{\nu}-p_{\nu}n_{\mu}\right)\psi_{\gamma}^{(t)}\left(u,q^{2}\right)+\left(e_{\mu}^{(\lambda=\perp)}n_{\nu}-e_{\nu}^{(\lambda=\perp)}n_{\mu}\right)h_{\gamma}^{(t)}\left(u,q^{2}\right)\right],\label{eq:VWF-munu}\end{eqnarray}

\begin{eqnarray}
\left\langle 0\left|\bar{\psi}\left(y\right)\psi\left(x\right)\right|V(q)\right\rangle  & = & f_{A}^{\perp}m_{A}^{2}e^{(\lambda)}\cdot z\int_{0}^{1}du\, e^{i(up\cdot y+\bar{u}p\cdot x)}\mathcal{D}_{T}\left(u,q^{2}\right),\label{eq:VWF}\end{eqnarray}
 \end{widetext} where the distribution amplitudes $\phi_{||},\phi_{\perp\gamma}$
have twist 2, $\psi_{\perp\gamma}^{(v)},\psi_{\gamma}^{(a)},\psi_{\gamma}^{(t)};\mathcal{D}_{T}$
has twist 3; and $h_{\gamma}^{(v)},h_{\gamma}^{(t)}$ have twist 4.
The formfactors $f_{\gamma}^{a}(q),f_{\perp\gamma}^{v}(q),f_{\perp\gamma}^{t}\left(q^{2}\right)$
and normalization constants $f_{3\gamma},\chi_{m},f_{A}^{\perp}$
are discussed in detail in~\cite{Dorokhov:2006qm}. 

 \end{document}